\documentclass[aps,prd,superscriptaddress,nofootinbib, 
preprintnumbers,floatfix,tightenlines,fleqn,notitlepage,10pt]{revtex4-1}
\usepackage{amsmath,amsfonts,amssymb,amscd,amsxtra,amsthm}
\usepackage{graphicx}  
\usepackage{epstopdf}
\usepackage{dcolumn}  
\usepackage{bm}          
\usepackage{slashed}
\usepackage{cancel}
\usepackage{float}
\usepackage{mathtools} 
\usepackage{amsbsy}
\usepackage{amstext}

\usepackage[utf8]{inputenc} 
\usepackage{booktabs} 
\usepackage[normalem]{ulem} 
\usepackage[dvipsnames]{xcolor} 
\usepackage{tabularx}
\usepackage{enumitem}  
\usepackage{array} 
\usepackage{slashed}
\usepackage{tikz}
\usepackage{float}
\usepackage{multirow}
\usepackage{cancel} 
\usepackage{physics}
\usepackage{hyperref}

\renewcommand\sout{\bgroup \color{red} \ULdepth=-.5ex \ULset}

\begin{document}
\preprint{INHA-NTG-06/2023}
\title{$D_{s0}^*(2317)$ and $B_{s0}^*$ as molecular states}
\author{Hee-Jin Kim}
\affiliation{Department of Physics, Inha University, Incheon 22212,
Republic of Korea}
\author{Hyun-Chul Kim}
\affiliation{Department of Physics, Inha University, Incheon 22212,
Republic of Korea}
\affiliation{School of Physics, Korea Institute for Advanced Study
(KIAS), Seoul 02455, Republic of Korea}
\date{\today}
\begin{abstract}
We investigate the dynamical generation of the $D_{s0}^*(2317)$ and
$B_{s0}^*$ mesons using a meson-exchange model with a coupled-channel
formalism. Our primary focus is on the $D_s^+\pi^0$ channel below the
$DK$ threshold. First, we construct the invariant kernel amplitudes,
incorporating effective Lagrangians based on heavy-quark symmetry,
flavor SU(3) symmetry, and hidden local symmetry. Since the
$D_{s0}^*(2317)$ state implies isospin symmetry breaking, we introduce
$\pi^0-\eta$ isospin mixing. We subsequently solve the coupled-channel
integral equations, which include four different channels, i.e., 
$D_s^+\pi^0$, $D^0 K^+$, $D^+ K^0$, and $D_s^+\eta$. We
carefully analyze how the pole corresponding to the $D_{s0}^*(2317)$
state emerges from the coupled channels. Our findings reveal that the
pole positions of the $D_{s0}^*(2317)$ meson are at $\sqrt{s_R}=(2317.9
- i 0.0593)$ MeV and the $\bar{B}_{s0}^*$ meson at $(5756.43-i0.0215)$
MeV, respectively. We also discuss their decay widths and
destructive interference of the two sources. In conclusion,
our current results provide a clear indication supporting the
interpretation of the $D_{s0}^*(2317)$ meson as a $DK$ molecular state 
within the present coupled-channel formalism. In addition, we discuss
a possible existence of $\bar{B}_{s0}^{*}$.
\end{abstract}
\maketitle

\section{Introduction} 
The $D_{s0}^*(2317)$ meson was initially discovered in 2003 by the
BaBar Collaboration while analyzing the inclusive $D_s^+\pi^0$
invariant mass distribution from $e^+e^-$ annihilation
data~\cite{BaBar:2003oey}. It was assigned a spin-parity of
$J^P=0^+$. Subsequently, both the CLEO and Belle Collaborations
confirmed the existence of this state~\cite{CLEO:2003ggt,
  Belle:2003guh}. Recently, the BESIII Collaboration made an
announcement that rekindled interest in the structure of the
$D_{s0}^*(2317)$. They observed it with a statistical
significance of $5.8\sigma$ and determined its mass to be $(2318.3\pm
1.2 \pm 1.2)$ MeV. Furthermore, they measured, for the first time, the
absolute branching ratio of $D_{s0}^*(2317)^{\pm}\to D_s^\pm \pi^0$
and found it to be $1.00_{-0.14}^{+0.00}
\mbox{(stat)}{}{-0.14}^{+0.00} \mbox{(syst)}$~\cite{BESIII:2017vdm}.   
Initially, it was assumed that the $D_{s0}^*(2317)$ meson is a
conventional $P$-wave $c\bar{s}$ state with $J^P=0^+$
\cite{Godfrey:1985xj, Godfrey:1986wj}. However, the 
observed mass based on this assumption turned out to be
approximately $150$ MeV lower than the results from quark models and
lattice calculations
\cite{Godfrey:1985xj, Godfrey:1986wj, Zeng:1994vj, Ebert:1997nk,
  Kalashnikova:2001ig, DiPierro:2001dwf, Dougall:2003hv}. This
discrepancy led to suggestions that the $D_{s0}^*(2317)$ meson could be
a non-$Q\bar{q}$ state, such as a molecular or tetraquark state
\cite{Barnes:2003dj, Szczepaniak:2003vy, Cheng:2003kg, Bali:2003jv,
  Hofmann:2003je, Kolomeitsev:2003ac, Guo:2006fu, Gamermann:2006nm,
  Lutz:2007sk, Faessler:2007gv, Xie:2010zza, Cleven:2010aw,
  Mohler:2013rwa, Liu:2012zya, Fu:2021wde}. 

The $D_{s0}^*(2317)$ lies in the proximity of the $DK$ threshold,
being located at about 45 MeV below it. This suggests the possibility
of considering it as a $DK$ molecular state. Furthermore, the decay
width of the $D_{s0}^*(2317)$ meson is observed to be extremely 
narrow, with only an upper limit known: $\Gamma < 3.8$ MeV at
the $95\%$ confidence level \cite{ParticleDataGroup:2022pth}. This
narrow decay width can be attributed to the strong isospin violation
involved in the $D_{s0}^*(2317)\to D_s\pi$ decay. Several theoretical
approaches have been proposed to explain the decay modes of
$D_{s0}^*(2317)$ \cite{Godfrey:2003kg, Colangelo:2003vg,
  Faessler:2007gv, Cleven:2014oka, Liu:2012zya,
  MartinezTorres:2014kpc, Song:2015nia, Fu:2021wde}. 
Given that the $D_{s0}^*(2317)$ is just below the $DK$
threshold, it is reasonable to investigate its nature as a dynamically
generated $DK$ molecular state. Several theoretical works have
provided strong support for this hypothesis. For instance, in
Refs.~\cite{Hofmann:2003je, Kolomeitsev:2003ac, Guo:2006fu,
  Gamermann:2006nm, Lutz:2007sk, Guo:2015dha,
  Fu:2021wde, Ikeno:2023ojl}, 
it was shown through the coupled-channel formalism that the
$D_{s0}^*(2317)$ can indeed be dynamically generated. These works
considered the $D_s\eta$ channel, taking into account the $\pi^0-\eta$
mixing. Lattice calculations have also considered the coupled-channel
effects. For example, Mohler et al.~\cite{Mohler:2013rwa} considered
both the $c\bar{s}$ and $DK$ interpolators, while in
Refs.~\cite{Liu:2012zya, MartinezTorres:2014kpc, Bali:2017pdv,
  Du:2017ttu, Gil-Dominguez:2023huq} the $DK$ and $D_s\eta$ channels
were introduced. For a comprehensive review of the experimental and 
theoretical status, we refer to the work by Chen et
al.~\cite{Chen:2016spr}.  

In this study, we aim to demonstrate how the $D_{s0}^*(2317)$ meson can
be dynamically generated as a $DK$ molecular state using the
coupled-channel formalism within the framework of the meson-exchange
model~\cite{Janssen:1994wn}. Janssen et al.~\cite{Janssen:1994wn}
previously showed the dynamical generation of the $f_0(980)$ and
$a_0(980)$ scalar mesons by constructing the $\pi\pi$, $\pi\eta$, and
$K\bar{K}$ coupled channels. Recently, a meson-exchange model with a
fully off-shell coupled-channel formalism was further developed and
successfully applied to describe the dynamically generated
axial-vector mesons $a_1(1260)$ and $b_1(1235)$ in $\pi\rho$
scattering~\cite{Clymton:2022jmv, Clymton:2023txd}. Considering that
the $f_0(980)$ and $a_0(980)$ mesons lie below the $K\bar{K}$
threshold and the masses of $a_1(1260)$ and $b_1(1235)$ are located
below the $K\bar{K}^{*}$ threshold, we expect that the
$D_{s0}^*(2317)$ meson can also be produced as a dynamically generated
scalar meson located below the $DK$ threshold. Therefore, we will
show in the current work that the $D_{s0}^*(2317)$ meson emerges as a
$DK$ molecular state within the framework of the meson-exchange
model. To achieve this, we compute the kernel amplitudes using
heavy-quark effective Lagrangians satisfying heavy-quark
symmetry~\cite{Isgur:1989vq, Georgi:1990um, Yan:1992gz}, flavor SU(3) 
symmetry, and hidden local symmetry~\cite{Bando:1984ej,
  Bando:1985rf, Bando:1987br}. We then solve the fully off-shell
coupled Blankenbecler-Sugar (BbS) integral equations derived from the
Bethe-Salpeter equation with three-dimensional
reduction~\cite{Blankenbecler:1965gx,  Aaron:1968aoz} with
four different channels considered.   

The structure of the present work is outlined as follows:
In Section II, we provide an overview of the general formalism
employed in the meson-exchange model with the coupled-channel
formalism.  In Section III, we demonstrate the emergence of the
$D_{s0}^*(2317)$ meson as a $DK$ molecular state and present the
corresponding numerical results. Additionally, we extend our analysis
to predict the existence of the $B_{s0}^*$ meson as a $BK$ molecular
state.  Finally, in the last section, we present our conclusions,
summarizing the key findings and contributions of this study.  
 
\section{Formalism} 
We first construct the the kernel amplitudes $V_{ij}$ for the
coupled-channel integral equation, where the subscripts $i$ and $j$
denote the possible transitions for the heavy mesons ($H$) and the
light pseudoscalar mesons ($\mathcal{M}$). Since we deal with the heavy
mesons, light pseudoscalar and vector mesons, we consider the relevant
symmetries for the effective Lagrangians. The heavy-quark spin-flavor
symmetry allows one to treat the $D$ and $B$ mesons on an equal
footing. The $\mathcal{M}$ arise as the pseudo-Nambu-Goldstone (pNG)
bosons from $\mathrm{SU}(3)_L\times \mathrm{SU(3)}_R$ chiral symmetry.  
Since we compute the transition amplitudes for the $H$ and
$\mathcal{M}$, we employ the effective chiral Lagrangian based on
heavy quark effective field 
theory~\cite{Wise:1992hn, Yan:1992gz, Casalbuoni:1996pg, Wise:1993wa}: 
\begin{align} 
\label{eq:1}
\mathcal{L}_\mathrm{heavy} = ig\mathrm{Tr}
[H_b \gamma_\mu \gamma_5 \mathcal{A}_{ba}^\mu \bar{H}_a]
+ i\beta\mathrm{Tr}
[H_b v^\mu (\mathcal{V}_\mu - \rho_\mu)_{ba} \bar{H}_a]
+ i\lambda\mathrm{Tr}[H_b\sigma^{\mu\nu}F_{\mu\nu}
(\rho)\bar{H}_a]
+ g_\sigma\bar{H}_aH_a\sigma,
\end{align}
where $H$ and $\bar{H}$ denote the $4\times 4$ superfields 
for the heavy pseudoscalar and vector meson fields $P$ and $P^*$ given 
in the the negative parity doublet  
\begin{align} 
\label{eq:2}
H^a = \frac{1+\slashed{v}}{2}(P_\mu^
{*a}\gamma^\mu-P^a\gamma_5), \;\;\;
\bar{H} = \gamma_0 H^\dagger \gamma_0
= (P_\mu^{*\dagger a}\gamma^\mu+P^{\dagger a}\gamma_5)
\frac{1+\slashed{v}}{2}.
\end{align}
$\mathcal{A}^\mu$ and $\mathcal{V}^\mu$ stand for the axial-vector
and vector currents consisting of the light pseudoscalar meson fields
expressed as   
\begin{align} 
\mathcal{A}^\mu &= \frac1{2}(\xi^\dagger\partial^\mu\xi -
\xi\partial^\mu\xi^\dagger), \cr
\mathcal{V}^\mu &= \frac1{2}(\xi^\dagger\partial^\mu\xi +
\xi\partial^\mu\xi^\dagger),
\label{eq:3}
\end{align}
where $\xi = e^{i\mathcal{M}/f_\pi}$ denotes the chiral field with the
pion decay constant $f_\pi=132$ MeV taken as a normalization
factor. The axial-vector current can be expanded in terms of the
pseudoscalar meson field 
\begin{align} 
\label{eq:4}
\mathcal{A}^\mu= \frac1{2}(\xi^\dagger\partial^\mu\xi -
\xi\partial^\mu\xi^\dagger) = \frac{i}{f_\pi}\partial^\mu \mathcal{M}
+\cdots, 
\end{align}
where the $\mathcal{M}$ represents the light pseudoscalar meson
matrix 
\begin{align} 
\mathcal{M} = \left(
\begin{array}{ccc}
\frac{\pi^0}{\sqrt{2}}+\frac{\eta}{\sqrt{6}} & \pi^+ & K^+
\\ 
\pi^- & -\frac{\pi^0}{\sqrt{2}}+\frac{\eta}{\sqrt{6}} & K^0
\\
K^- & \bar{K}^0 & -\sqrt{\frac{2}{3}}\eta
\end{array}\right).
\label{eq:5}
\end{align}
The vector meson octet is introduced as the gauge
field of the hidden local symmetry~\cite{Bando:1984ej, Bando:1985rf}
\begin{align} 
\rho^\mu = i\frac{g_V}{\sqrt{2}} V^\mu,\;\;\;
V^\mu=\left(\begin{array}{ccc}
\frac{\rho^0}{\sqrt{2}}+\frac{\omega}{\sqrt{2}} & \rho^+ & K^{*+}
\\ 
\rho^- & -\frac{\rho^0}{\sqrt{2}}+\frac{\omega}{\sqrt{2}} & K^{*0}
\\
K^{*-} & \bar{K}^{*0} & \phi
\end{array}\right)^\mu.
\label{eq:6}
\end{align}
The field strength tensor is defined as
$F^{\mu\nu}=\partial^\mu\rho^\nu - \partial^\nu\rho^\mu +  
[\rho^\mu,\rho^\nu]$ with $\rho^\mu = \rho^{\mu a}\lambda^a$. 

The first term of Eq.~\eqref{eq:1} is reduced to the effective
Lagrangians for the $PP^*M$ and $P^*P^*M$ vertices
\begin{align} 
\mathcal{L}_{PP^*M} &= -\frac{2g}{f_\pi} P_b^{*\mu}
\partial_\mu \mathcal{M}_{ab} P_a^\dagger + \mathrm{h.c.},
\cr
\mathcal{L}_{P^*P^*M} &= \frac{2ig}{f_\pi} P_b^{*\beta}
\partial^\mu \mathcal{M}_{ab} v^\nu P_a^
{*\alpha\dagger} \varepsilon_{\alpha\beta\mu\nu} + 
\mathrm{h.c.},
\label{eq:7}
\end{align}
where $P$ is represented as the heavy meson antitrplet in flavor
SU(3), i.e., $(D^0,\,D^+, D_s)$ for $Q=c$, whereas it is written as
$(B^-, \, B^0,\,B_s)$ for $Q=b$. 

The axial coupling $g=0.59$ is extracted from the $D^*$ decay
width~\cite{CLEO:2001foe}. $\beta=0.9$ is determined by the vector
meson dominance~\cite{Bando:1987br}. The parameter $\lambda$ in the
effective Lagrangian given in Eq.~\eqref{eq:1} is related to the
coupling constants for the $PP^*V$ vertex and the tensor coupling
constants for the $P^*P^*V$ vertex. In Refs.~\cite{Casalbuoni:1996pg,
  Isola:2003fh}, the value of $\lambda$ was estimated in the following
way: the phenomenological heavy-to-light current was constructed, from
which the heavy-light vector and axial-vector transition form factors
were examined in a similar manner that the heavy-to-heavy matrix
elements can be parametrized in terms of the 
Isgur-Wise form factor~\cite{Isgur:1989vq,Wise:1992hn}.
The heavy-to-light current~\cite{Casalbuoni:1996pg, Isola:2003fh} is
expressed as  
\begin{align}
L_a^{\mu} = \frac{i\hat{F}}{2}\mathrm{Tr}\left(
 \gamma^\mu(1-\gamma^5) H_b \xi_{ba}^\dagger\right).
\end{align}
The coefficient $\hat{F}$ is related to the leptonic decay
constant of the pseudoscalar heavy meson, $f_B$, which was estimated
in QCD sum rules~\cite{Casalbuoni:1996pg}. The value of $\lambda$ can
be related to the $B\to K^*$ vector transition form factor at high
$q^2$ within the effective field theory approach. With the results
derived from the light-cone QCD sum rules and lattice QCD matched,  
the value of $\lambda$ was determined to be $\lambda=0.56\,
\mathrm{GeV}^{-1}$. The last term in Eq.~\eqref{eq:1} provides the
interaction with the $\sigma$ meson and its coupling $g_\sigma$ will
be discussed later.   

The effective Lagrangians for the $PPV$, $PP^*V$, and $P^*P^*V$
vertices can be derived from the second and third terms of
Eq.~\eqref{eq:1}:   
\begin{align} 
\mathcal{L}_{PPV} &= -2i\beta PP^\dagger v\cdot
\rho = -\sqrt{2}\beta g_V PP^\dagger v\cdot V, \cr
\mathcal{L}_{PP^*V} &= -4i \lambda \varepsilon_
{\mu\nu\alpha\beta} (P\partial^\mu
P^{*\dagger\nu}+P^\dagger\partial^\mu P^{*\nu})v^\alpha
\rho^\beta \cr
&= 2\sqrt{2} \lambda g_V \varepsilon_{\mu\nu\alpha\beta} 
(P\partial^\mu P^{*\dagger\nu}+P^\dagger\partial^\mu P^
{*\nu})v^\alpha V^\beta ,\cr
\mathcal{L}_{P^*P^*V} &= -2i\beta P^{*\mu} P_\mu^
{*\dagger} v\cdot \rho - 4\lambda (\partial_\mu
P^{*\dagger\mu} P_\nu^* - \partial_\mu P^{*\mu} 
P_\nu^{*\dagger})\rho^\nu \cr
&= \sqrt{2}\beta g_V P^{*\mu} P_\mu^
{*\dagger} v\cdot V - 2\sqrt{2}i\lambda g_V
(\partial_\mu P^{*\dagger\mu} P_\nu^* - \partial_\mu P^
{*\mu} P_\nu^{*\dagger}) V^\nu,
\label{eq:8}
\end{align}
where $g_V=5.8$~\cite{Bando:1987br}. To compute the kernel amplitudes
$V_{ij}$, we extract the effective Lagrangians from Eqs.~\eqref{eq:1}
and \eqref{eq:8} 
\begin{align} 
\mathcal{L}_{DD^*\mathcal{M}} &= -g_{DD^*\mathcal{M}} (D^
{*\mu} D^\dagger + D D^{*\mu\dagger})\partial_\mu
\mathcal{M}, \cr
\mathcal{L}_{DDV} &= ig_{DDV} (D\partial_\mu D^\dagger
-\partial_\mu D^\dagger D)V^\mu, \cr
\mathcal{L}_{DD\sigma} &= 2g_{DD\sigma} m_D DD\sigma,
\label{eq:9}
\end{align}
where the value of the $DD^*\mathcal{M}$ coupling constant is then
rescaled as $g_{DD^*\mathcal{M}}=2g\bar{m}_D/f_\pi=17.6$. Similarly, 
the $g_{BB^*\mathcal{M}}$ can be evaluated as $g_{BB^*\mathcal{M}} =
2g \bar{m}_B/f_\pi = 49.0$. We have used the average mass of
neutral and charged $D(B)$ mesons, $\bar{m}_{D(B)}$ for rescaling.  

It is of great difficulty to determine the $PP\sigma$ and $PP\rho$
coupling constants empirically and experimentally. To fix the value
of $g_{PP\sigma}$, Refs.~\cite{Liu:2009qhy, Liu:2010xh} employed the
relation $g_\sigma=g_\pi/2\sqrt{6}=0.76$, assuming that $\sigma$ is
the chiral partner of the pion also in the heavy-light quark
systems~\cite{Bardeen:2003kt}. As pointed out in Ref.~
\cite{Liu:2009qhy}, since the $\sigma$ meson has a very broad
width, its mass cannot be determined uniquely. Moreover, the effects
of its width should also be considered in the coupling constant.  
In Ref.~\cite{Kim:2019rud}, the $g_{DD\sigma}$ and $g_{BB\sigma}$ were
determined by using the same coupled-channel formalism as the current
work, and the dispersion relation. We first constructed the
rescattering equation for $D\bar{D}\to \pi\pi$ that contains $\pi\pi$
$T$ matrix. Having projected the $T_{D\bar{D}\to \pi\pi}$ into the
scalar ($J=0$) and isoscalar ($I=0$) S-wave channel, we evaluated the
spectral function for $T_{D\bar{D}\to D\bar{D}}^{J=I=0}$ partial-wave
matrix element. Inserting it into the dispersion relation, we were
able to extract $g_{DD\sigma}$. We can also obtain $g_{DD\rho}$ in the
same manner. Similarly, the $\sigma$ and $\rho$ coupling constants for
the $B$ meson were also obtained. The results are given below
\begin{align} \label{}
g_{DD\sigma} &= 1.50, \;\;\; g_{DD\rho} = 1.65, \cr
g_{BB\sigma} &= 7.05, \;\;\; g_{BB\rho} = 8.92.
\end{align}
The large values of the coupling constants for the $B$ meson arise
from the mass prefactor. 

Since we employ the meson-exchange picture to calculate $V_{ij}$, we
take the effective Lagrangians for the $\mathcal{M}\mathcal{M}V$ from
the hidden local symmetry~\cite{Bando:1984ej, Bando:1985rf, Bando:1987br}. 
It is known that chiral symmetry $\mathrm{SU(3)}_L \otimes
\mathrm{SU(3)}_R$ is spontaneously broken to the vector subgroup 
$\mathrm{SU(3)}_V$. Consequently, the pNG bosons emerge from the coset
space $\mathrm{SU(3)}_L \otimes
\mathrm{SU(3)}_R/\mathrm{SU(3)}_V$. The effective chiral Lagrangian  
for the NG bosons was constructed in the coset
space~\cite{Callan:1969sn, Gasser:1983yg}.      
Bando et al.~\cite{Bando:1984ej, Bando:1985rf} proposed the gauge
equivalence between $\mathrm{SU(3)}_L \otimes
\mathrm{SU(3)}_R/\mathrm{SU(3)}_V$ and $[\mathrm{SU(3)}_L \otimes 
\mathrm{SU(3)}_R]\otimes [\mathrm{SU(3)}_V]_{\mathrm{local}}$, where
$[\mathrm{SU(3)}_V]_{\mathrm{local}}$ stands for the hidden local
symmetry and the pertinent gauge bosons appear as the composite
fields. Bando et al.~\cite{Bando:1984ej} showed that
the kinetic term of the nonlinear effective chiral Lagrangian can be 
expressed as the linear effective Lagrangian in terms of the gauge
vector and and pNG bosons. In Ref.~\cite{Bando:1985rf}.
The dynamical gauge vector bosons from the hidden local symmetry are
identified as  $\rho$, $\omega$, and $K^*$. We take the effective
Lagrangian for the $\mathcal{M}\mathcal{M}V$ vertices from the hidden
local symmetry   
 \begin{align}
   \mathcal{L}_{\mathcal{M}\mathcal{M}V} &= -\frac{i}{2}
  g_{\mathcal{M}\mathcal{M}V}\,\mathrm{Tr}([\mathcal{M}^\dagger,\partial_\mu  
\mathcal{M}]V_\mu) 
\end{align}
with $g_{\mathcal{M}\mathcal{M}V} = g_{\pi\pi\rho}$ and
$g_{\pi\pi\rho}^2/4 \pi=2.84$. 
The effective Lagrangian for the $\mathcal{M}\mathcal{M}\sigma$
vertices can be constructed as 
\begin{align}
\mathcal{L}_{\mathcal{M}\mathcal{M}\sigma} &= 2g_{\mathcal{M}
\mathcal{M}\sigma} m_\mathcal{M} \mathrm{Tr}(\mathcal{M}^\dagger 
\mathcal{M})\sigma,
\end{align}
where $g_{\pi^0\pi^0\sigma} = 8.7$ is extracted from the S-wave
$\pi\pi$ scattering amplitude. We fix the value of $g_{\eta\eta\sigma}=5.0$
which is similar with the $g_{\pi^0\pi^0\sigma}$ due to the lack of
experimental data. The SU(3) relations for the coupling constants are
given in Appendix.~\ref{app:a1}. 

Since $D_{s0}^*(2317)$ lies below the $DK$ threshold, it is forbidden
to decay $DK$ kinematically. It is remarkable that $D_{s0}^*(2317)$
decays into $D_s\pi^0$, which requires the breakdown of isospin
symmetry. While the upper limit of the decay width $\Gamma_{D_s\pi}$
is only known, we expect that $\Gamma_{D_s\pi}$ should be
tiny because the effects of isospin symmetry breaking originates from
the mass difference of the current up and down quarks and
electromagnetic interaction.  We only consider the mass difference
between $m_u$ and $m_d$ as a main source of isospin symmetry
breaking. Then, the neutral pion can be mixed with the $\eta$ meson,
so that the effective Lagrangian for the $\pi^0-\eta$ mixing can be
expressed as~\cite{Gross:1979ur, Cho:1994zu, Gasser:1984gg}:  
\begin{align} 
\label{}
\mathcal{L}_{\pi\eta} = -\frac{B_0}{\sqrt{3}}
(m_u-m_d)\pi^0\eta.
\end{align}
Introducing the mixing angle $\epsilon$, we can construct the 
$|D_s^+\pi^0\rangle$ state with the $D_s^+ \eta$ component 
\begin{align} 
\left(\begin{array}{c}
\pi^0 \\ \eta
\end{array}\right) = 
\left(\begin{array}{cc}
\cos\epsilon & -\sin\epsilon \\
\sin\epsilon & \cos\epsilon
\end{array}\right)
\left(\begin{array}{c}
\tilde{\pi}^0 \\ \tilde{\eta}
\end{array}\right),
\label{eq:mixing}
\end{align}
The value of the mixing angle can be estimated by using the 
up and down current quark masses~\cite{ParticleDataGroup:2022pth}  
\begin{align} \label{}
\tan 2\epsilon = \frac{\sqrt{3}}{2} 
\frac{m_d-m_u}{m_s - (m_u+m_d)/2}, \;\;\;
\epsilon = 0.012.
\end{align} 
The $D_s^+\pi^0$ and $D_s^+\eta$ states are then expressed in
terms of the decoupled fields $\tilde{\pi}^0$ and $
\tilde{\eta}$
\begin{align} \label{}
&|D_s^+\pi^0\rangle = |D_s^+\tilde{\pi}^0\rangle
\cos\epsilon
- |D_s^+\tilde{\eta}\rangle \sin\epsilon, \cr
&|D_s^+\eta\rangle = |D_s^+\tilde{\pi}^0\rangle
\sin\epsilon + |D_s^+\tilde{\eta}\rangle \cos\epsilon.
\end{align}

\begin{figure}[H]
\centering
\includegraphics[scale=0.8]{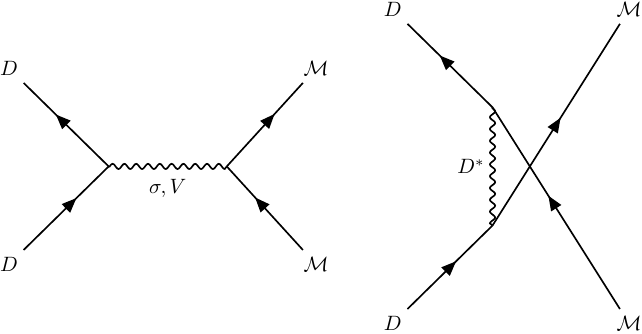}
\caption{Feynman diagrams for the $t$ and $u$ channels. $D$
and $\mathcal{M}$ stand for the external charmed and light
psuedoscalar mesons, respectively. $\sigma$ and light vector mesons
are exchanged in the $t$ channel, whereas the vector charmed mesons
are exchanged in the $u$ channel.}
\label{fig:1}
\end{figure}

We now compute the kernel matrix for the two-body
processes with charm and strangeness $+1$. Considering the
isospin breaking channel $D_s^+\pi^0$, we include the 
four different channels in the charge basis: $D_s^+\pi^0$,
$D^0K^+$, $D^+K^0$ and $D_s^+\eta$. Therefore, $\mathcal{V}$ is
constructed as   
\begin{align} \label{}
\mathcal{V} = \left(\begin{array}{cccc}
\mathcal{V}_{D_s^+\pi^0\to D_s^+\pi^0} & 
\mathcal{V}_{D_s^+\pi^0\to D^0K^+} &
\mathcal{V}_{D_s^+\pi^0\to D^+K^0} &
\mathcal{V}_{D_s^+\pi^0\to D_s^+\eta} \\
\mathcal{V}_{D^0K^+\to D_s^+\pi^0} & 
\mathcal{V}_{D^0K^+\to D^0K^+} &
\mathcal{V}_{D^0K^+\to D^+K^0} &
\mathcal{V}_{D^0K^+\to D_s^+\eta} \\
\mathcal{V}_{D^+K^0\to D_s^+\pi^0} & 
\mathcal{V}_{D^+K^0\to D^0K^+} &
\mathcal{V}_{D^+K^0\to D^+K^0} &
\mathcal{V}_{D^+K^0\to D_s^+\eta} \\
\mathcal{V}_{D_s^+\eta\to D_s^+\pi^0} & 
\mathcal{V}_{D_s^+\eta\to D^0K^+} &
\mathcal{V}_{D_s^+\eta\to D^+K^0} &
\mathcal{V}_{D_s^+\eta\to D_s^+\eta}
\end{array}
\right).
\end{align}
At the tree level, we derive the relevant Feynman diagrams, i.e.,
$\sigma$- and vector-meson exchanges in the $t$ channel, and vector 
charmed meson exchange in the $u$ channel as shown in 
Fig.~\ref{fig:1}.  
Using the effective Lagrangians given above, we obtain the Feynman
amplitudes as follows: 
\begin{align} 
\mathcal{V}_\sigma^t(t) &= 4F^2(t)g_{DD\sigma} g_
{\mathcal{M}\mathcal{M}\sigma}m_Dm_\mathcal{M} \frac{1}{t-m_\sigma^2},
\cr
\mathcal{V}_V^t(t) &= -F^2(t)g_{DDV} g_{\mathcal{M}\mathcal{M}V}
\frac{(p_1+k_1)\cdot(p_2+k_2)}{t-m_V^2},\cr
\mathcal{V}_{D^*}^u(u) &= F^2(u)g_{DD^*\mathcal{M}}^2
\frac{(p_1+k_2)\cdot(p_2+k_1)}{u-m_{D^*}^2},
\label{}
\end{align}
where $p_i$ and $k_i$ denote the inital and final four-momenta,
respectively. The Mandelstam variables are defined as
\begin{align} \label{eq:18}
t = (p_1-k_1)^2 = (p_2-k_2)^2, \;\;\;
u = (p_1-k_2)^2 = (p_2-k_1)^2.
\end{align}
We want to emphasize that any pole diagrams in the $s$ channel are
excluded, because we will explicitly show that the $D_{s0}^*$ and
$B_{s0}^*$ are dynamically generated. The values of the masses for the
mesons involved in the current work are listed in Table~\ref{tab:1}, 
sourced from PDG data~\cite{ParticleDataGroup:2022pth}. Note that
since the effects of isospin symmetry breaking are essential in the
production of $D_{s0}^*$, we consider the isospin mass differences for
the mesons involved. 

\setlength{\tabcolsep}{15pt}
\renewcommand{\arraystretch}{1.5}
\begin{table}[htp]
\caption{Particle masses and spin-parity quantum numbers.}
\label{tab:1}
\begin{tabular}{lcc|lcc}
\hline
\hline
Particle & Mass[MeV] & $J^{PC}$ & Particle & Mass[MeV] & $J^{PC}$ \\
\hline 
$\pi^0$ & $134.98$ & $0^{-+}$ & $D^0(\bar{D}^0)$ & $1864.84$ & $0^-$
\\
$\eta$ & $547.86$ & $0^{-+}$ & $D^\pm$ & $1869.66$ & $0^-$ \\
$\sigma$ & $550$ & $0^{++}$ & $D^{*0}(\bar{D}^{*0})$ & $2006.85$ &
$1^-$
\\
$\rho$ & $770$ & $1^{--}$ & $D^{*\pm}$ & $2010.26$ & $1^-$ \\
$\omega$ & $782.66$ & $1^{--}$ & $D_s^\pm$ & $1968.35$ & $0^-$ \\
$K^0(\bar{K}^0)$ & $497.611$ & $0^-$ & $D_s^{*\pm}$ & $2112.2$ & $1^-$
\\
$K^\pm$ & $493.677$ & $0^-$ &&&\\
$K^{*0}(\bar{K}^{*0})$ & $895.55$ & $1^-$ &&&\\
$K^{*\pm}$ & $891.67$ & $1^-$ &&&\\
\hline
\hline
\end{tabular}
\end{table}

Since the hadrons have finite sizes, we have to introduce the form
factor at each vertex. We choose the following type of the form factors: 
\begin{align} \label{}
F(q^2) = \left(\frac{n\Lambda^2-m_\mathrm{ex}^2}
{n\Lambda^2-q^2}\right)^n,
\end{align}
where $\Lambda$ stands for the cutoff mass and $m_\mathrm{ex}$
designates the mass of the exchange meson. The $q^2$ is 
the squared momentum transfer between incoming and outgoing states. 
While a specific value of $\Lambda$ depends on the size of a meson, 
we have no experimental and empirical information on it.
Values of the cutoff masses bring about most of the uncertainties in
any meson-exchange models because of lack of experimental information
on them. To reduce the uncertainties that may be caused by the cutoff 
masses, we utilize the idea that a heavier hadron is more compact than 
a light one~\cite{Kim:2018nqf, Kim:2021xpp, Won:2022cyy}, based on which
we introduce a prescription that the value of $\Lambda$ can be taken to be
$\Lambda \simeq (m_\mathrm{ex}+600)$ MeV. It was phenomenologically
successful in describing various hadronic processes~\cite{Kim:2019rud,
  Clymton:2022jmv, Clymton:2023txd}, in particular, when heavy hadrons
are involved~\cite{Clymton:2021thh, Kim:2019rud}. We emlpoy slightly
smaller values of $\Lambda$ ($\Lambda\simeq (m_{\mathrm{ex}}+540)$
MeV) for the $u$ channel diagrams for locating the proper pole
position. The dipole-type form factor $n=2$ is used in this work. We
list the cutoff parameters in Table~\ref{tab:2}. 
\begin{table}[htp]
\caption{The cutoff parameters in MeV for possible exchange
diagrams for each reaction.}
\label{tab:2}
\begin{ruledtabular}
\begin{tabular}{lccr}
Reaction  & Exchange & Type & $\Lambda-m_\mathrm{ex}$ [MeV]
\\
\hline 
$D_s^+\pi^0\to D_s^+\pi^0$ & $\sigma$ & $t$ & $600$ \\
$D_s^+\pi^0\to D_s^+\eta$ & $\sigma$ & $t$ & $600$ \\
\multirow[t]{2}{*}{$D_s^+\pi^0(\eta)\to D^0K^+$} 
& $K^{*+}$ & $t$ & $600$ \\
& $D^{*0}$ & $u$ & $530$ \\
\multirow[t]{2}{*}{$D_s^+\pi^0(\eta)\to D^+K^0$} 
& $K^{*0}$ & $t$ & $600$ \\
& $D^{*+}$ & $u$ & $530$ \\
\multirow[t]{3}{*}{$D^0K^+\to D^0K^+$}
& $\rho^0$ & $t$ & $600$ \\
& $\omega$ & $t$ & $600$ \\
& $D_s^{*-}$ & $u$ & $540$ \\
\multirow[t]{3}{*}{$D^+K^0\to D^+K^0$} 
& $\rho^0$ & $t$ & $600$ \\
& $\omega$ & $t$ & $600$ \\
& $D_s^{*+}$ & $u$ & $540$ \\
$D^0K^+\to D^+K^0$ & $\rho^-$ & $t$ & $600$ \\
\end{tabular}%
\end{ruledtabular}
\end{table}

The kernel matrix elements are then obtained as
follows: 
\begin{align} \label{}
\mathcal{V}_{D_s^+\pi^0\to D_s^+\pi^0} &= 
\mathcal{V}_{D_s^+\tilde{\pi}^0\to D_s^+\tilde{\pi}^0}^
{\sigma,t}\cos^2\epsilon + 
\mathcal{V}_{D_s^+\tilde{\eta}\to D_s^+\tilde{\eta}}^
{\sigma,t}\sin^2\epsilon, \cr
\mathcal{V}_{D_s^+\eta\to D_s^+\eta} &= 
\mathcal{V}_{D_s^+\tilde{\pi}^0\to D_s^+\tilde{\pi}^0}^
{\sigma,t}\sin^2\epsilon +
\mathcal{V}_{D_s^+\tilde{\eta}\to D_s^+\tilde{\eta}}^
{\sigma,t}\cos^2\epsilon, \cr
\mathcal{V}_{D_s^+\pi^0\to D_s^+\eta} &= \left(
\mathcal{V}_{D_s^+\tilde{\eta}\to D_s^+\tilde{\eta}}^
{\sigma,t} -
\mathcal{V}_{D_s^+\tilde{\pi}^0\to D_s^+\tilde{\pi}^0}^
{\sigma,t} \right) \sin\epsilon \cos\epsilon, \cr
\mathcal{V}_{D_s^+\pi^0\to D^0K^+} &=
\left(\mathcal{V}_{D_s^+\tilde{\pi}^0\to D^0K^+}^{K^
{*+},t} + \mathcal{V}_{D_s^+\tilde{\pi}^0\to D^0K^+}^{D^
{*0},u}\right)\cos\epsilon
- \left(\mathcal{V}_{D_s^+\tilde{\eta}\to D^0K^+}^{K^
{*+},t} + \mathcal{V}_{D_s^+\tilde{\eta}\to D^0K^+}^{D^
{*0},u}\right) \sin\epsilon, \cr
\mathcal{V}_{D_s^+\pi^0\to D^+K^0} &=
\left(\mathcal{V}_{D_s^+\tilde{\pi}^0\to D^+K^0}^{K^
{*0},t} + \mathcal{V}_{D_s^+\tilde{\pi}^0\to D^+K^0}^{D^
{*+},u}\right)\cos\epsilon
- \left(\mathcal{V}_{D_s^+\tilde{\eta}\to D^+K^0}^{K^
{*0},t} + \mathcal{V}_{D_s^+\tilde{\eta}\to D^+K^0}^{D^
{*+},u}\right) \sin\epsilon, \cr
\mathcal{V}_{D_s^+\eta\to D^0K^+} &=
\left(\mathcal{V}_{D_s^+\tilde{\pi}^0\to D^0K^+}^{K^
{*+},t} + \mathcal{V}_{D_s^+\tilde{\pi}^0\to D^0K^+}^{D^
{*0},u}\right)\sin\epsilon
+ \left(\mathcal{V}_{D_s^+\tilde{\eta}\to D^0K^+}^{K^
{*+},t} + \mathcal{V}_{D_s^+\tilde{\eta}\to D^0K^+}^{D^
{*0},u}\right) \cos\epsilon, \cr
\mathcal{V}_{D_s^+\eta\to D^+K^0} &=
\left(\mathcal{V}_{D_s^+\tilde{\pi}^0\to D^+K^0}^{K^
{*0},t} + \mathcal{V}_{D_s^+\tilde{\pi}^0\to D^+K^0}^{D^
{*+},u}\right)\sin\epsilon
+ \left(\mathcal{V}_{D_s^+\tilde{\eta}\to D^+K^0}^{K^
{*0},t} + \mathcal{V}_{D_s^+\tilde{\eta}\to D^+K^0}^{D^
{*+},u}\right) \cos\epsilon, \cr
\mathcal{V}_{D^0K^+\to D^0K^+} &=
\mathcal{V}_{D^0K^+\to D^+K^0}^{\rho^0,t}
+ \mathcal{V}_{D^0K^+\to D^+K^0}^{\omega,t} 
+ \mathcal{V}_{D^0K^+\to D^+K^0}^{D_s^{*-},u}, \cr
\mathcal{V}_{D^+K^0\to D^+K^0} &=
\mathcal{V}_{D^+K^0\to D^+K^0}^{\rho^0,t}
+ \mathcal{V}_{D^+K^0\to D^+K^0}^{\omega,t} 
+ \mathcal{V}_{D^+K^0\to D^+K^0}^{D_s^{*+},u}, \cr
\mathcal{V}_{D^0K^+\to D^+K^0} &=
\mathcal{V}_{D^0K^+\to D^+K^0}^{\rho^-,t}.
\end{align}
Since no neutral $D_s^*$ state exists, the $u$ channel diagrams for
$D^0K^+\to D^+K^0$ amplitude are absent. We
will discuss this effect later. The $D_s^+\pi^0$ channel provides a
background. The two $DK$ channels play key roles in generating the
$D_{s0}^*(2317)$ resonance, while the $D_s^+\eta$ channel shifts 
the pole position. 

\begin{figure}[htp]
    \centering
    \includegraphics[scale=.68]{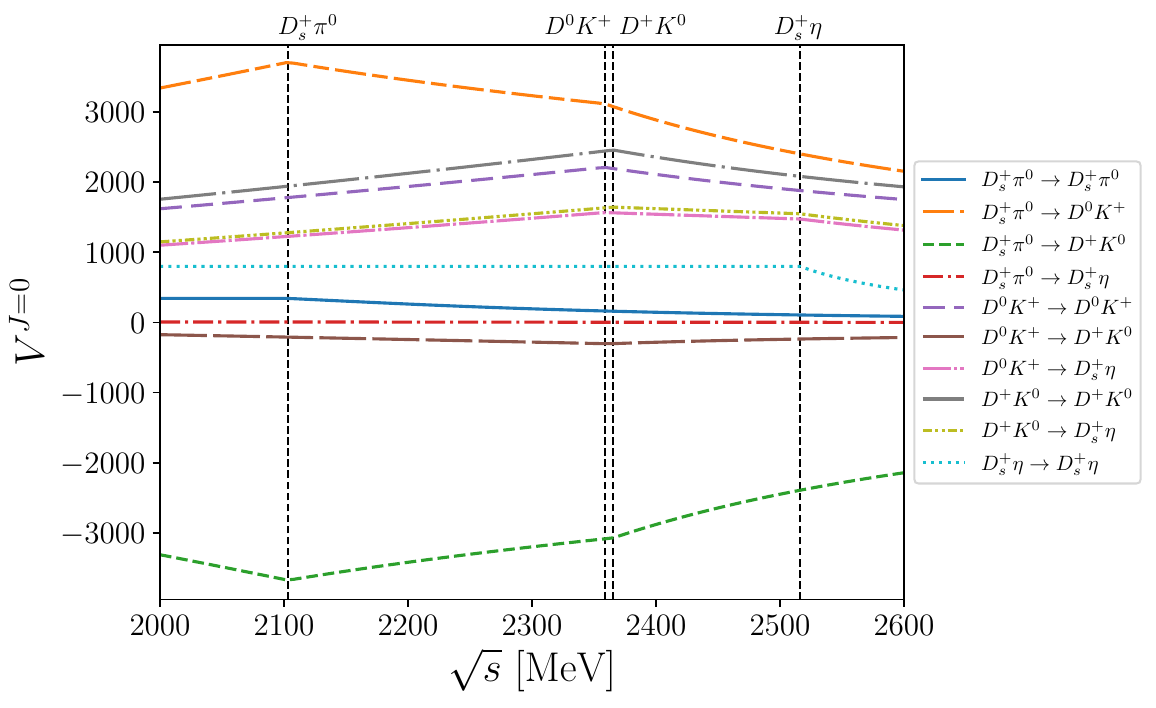}
    \caption{The partial-wave kernel amplitudes in four different
    channels. The dashed vertical lines represent the threshold
    energies for the channels involved in the current
    coupled-channel formalism.}
    \label{fig:2}
\end{figure}

The matrix elements of $\mathcal{V}$ are depicted in Fig.~\ref{fig:2}.  
We observe that the transition amplitudes $\mathcal{V}_{D_s^+\pi^0\to
  D^0K^+}$ and $\mathcal{V}_{D_s^+\pi^0\to D^+K^0}$ dominate over
all other contributions, primarily due to the large coupling constants 
in the $D^*$ exchange diagrams in the $u$ channel. The $DK\to DK$
amplitudes provide significant contributions because of the large
value of $g_{DD^*\mathcal{M}}$ in the $u$ channel. The $D_s^+\eta \to
DK$ amplitude contains the SU(3) symmetric factor of $1/\sqrt{6}$,
which is smaller than that of $1/\sqrt{2}$ for the  $D_s^+\pi^0$
channel as shown in Appendix.~\ref{app:a1}. Consequently, it
gives a smaller contribution to the $T$-matrix. In the case of the  
$D^0K^+\to D^+K^0$ amplitudes, no $u$ channel process is allowed due 
to the absence of the neutral $D_s^*$. As a result, its contribution
is expected to be marginal.

We are now in a position to evaluate the coupled integral
equations. The scattering amplitude is defined by 
\begin{align} 
S_{fi} = \delta_{fi} - (2\pi)^4\delta^{(4)}
(P_f-P_i) T_{fi},
\label{eq:22}
\end{align}
where $P_i$ and $P_f$ denote the total four-momenta of the
initial and final states, respectively. 
We use the Blankenbecler-Sugar three-dimensional reduction for the
Bethe-Salpeter equations~\cite{Blankenbecler:1965gx, Aaron:1968aoz}: 
\begin{align} 
T_{fi}(p,p')=V_{fi}(p,p')+\frac 1{(2\pi)^4}
\int d^4q\, V_{fk}(p,q)G_k(q) T_{ki}(q,p').
\label{eq:23}
\end{align}
The propagator $G_k(q)$ takes the following form 
\begin{align} 
G_k(q) = \frac{\pi}{\omega_1^k \omega_2^k}
\delta\left(q^0 - \frac{\omega_1^k-\omega_2^k}
{2}\right) \frac{\omega_1^k+\omega_2^k}{
s-(\omega_1^k+\omega_2^k)^2+i\epsilon}.
\label{eq:24}
\end{align}
Defining the on-shell energies of particles 1 and 2 for the $k$
channel, $\omega_{1(2)}^k$, to be $\omega_{1(2)}^k
=\sqrt{\bm{q}^2+(m_{1(2)}^k)^2}$, we obtain the 
Blankenbecler-Sugar(BbS) equations: 
\begin{align} 
T_{fi}(\bm{p},\bm{p'})=V_{fi}(\bm{p},\bm{p'})+ \int d^3q 
\, V_{fk}(\bm{p},\bm{q}) \tilde{G}_k(\bm{q}) T_{ki}(\bm{q},
\bm{p'}),
\label{}
\end{align}
where
\begin{align} \label{}
\tilde{G}_k(\bm{q}) = \frac 1{(2\pi)^3}
\frac{\omega_1^k+\omega_2^k}{2\omega_1^k\omega_2^k}
\frac{1}{s-(\omega_1^k+\omega_2^k)^2+i\epsilon}.
\end{align}

To generate the scalar $D_{s0}^*(2317)$ meson dynamically, we only need 
the S-wave $T$ amplitude, $T^{J=0}$. So, we carry out the partial-wave
expansion of the BbS equation
\begin{align} 
T_{fi}^J(p,p') = V_{fi}^J(p,p') + \frac 1{
(2\pi)^3} \int_0^\infty dq\, 
\frac{\omega_1^k+\omega_2^k}{2\omega_1^k\omega_2^k}
\frac{q^2V_{fk}^J(p,q) T_{ki}^J(q,p')}{
s-(\omega_1^k+\omega_2^k)^2+i\epsilon},
\label{eq:27}
\end{align}
where $V^J$ and $T^J$ are the partial-wave kernel and transition 
amplitudes defined by 
\begin{align}
V^J =2\pi \int_{-1}^{1} dx V(p,k;x) P_J(x),\;\;\;
T^J =2\pi \int_{-1}^{1} dx T(p,k;x) P_J(x).
\end{align}
$P_J(x)$ stands for the Legendre polynomial. 
Since Eq.~\eqref{eq:27} contains a singulraity arising from the
two-body propgator, we need to isolate the singular part as follows 
\begin{align} \label{}
T_{fi}^J(p,p') &= V_{fi}^J(p,p') + \frac1{(2\pi)^3} \sum_k \left[
\int_0^\infty dq q\frac{\omega_1^k+\omega_2^k}{2\omega_1^k\omega_2^k}
\frac{qV_{fk}^J(p,q)T_{ki}^J(q,p') - \tilde{q}_k
V_{fk}^J(p,\tilde{q}_k)T_{ki}^J(\tilde{q}_k,p')}
{s-(\omega_1^k+\omega_2^k)^2} \right.\cr
&\qquad \left. + \frac1{2\sqrt{s}}\left(\log\left|\frac{
\sqrt{s}-\omega_1^k-\omega_2^k}
{\sqrt{s}+\omega_1^k+\omega_2^k}\right|-i\pi\right)\tilde{q}_k
V_{fk}^J(p,\tilde{q}_k)T_{ki}^J(\tilde{q}_k,p')
\right],
\end{align}
where the second term in the bracket is ths singular part.
$\tilde{q}_k$ denotes the momentum point, where the singularity arises
in the denominator. Constructing the off-shell partial-wave $V^J$ and
solving the partial-wave integral equation, we can derive the
partial-wave $T^J$ matrix elements. 
Formally, the integral equation can be solved by using the matrix
inversion: 
\begin{align} \label{}
T^J=(1-V^J\tilde{G})^{-1}V^J.
\end{align}
Poles corresponding to resonance states can be found by
evaluating the corresponding zeros of the following equation
$\mathrm{det}(1-V^J(s)\tilde{G}^J(s))=0$. 

\section{Results and Discussion}
We first examine the transition amplitudes $T^{J=0}$ in the complex
plane to see how the poles appear. To analyze the coupled-channel
effects, we strictly follow our cutoff scheme $\Lambda =
(m_\mathrm{ex}+600)$ MeV for a while. Note that the kernel amplitudes do
not contain any singularities because we do not include the $s$ channel
diagrams. 
\begin{figure}[htp]
    \centering
    \includegraphics[scale=.44]{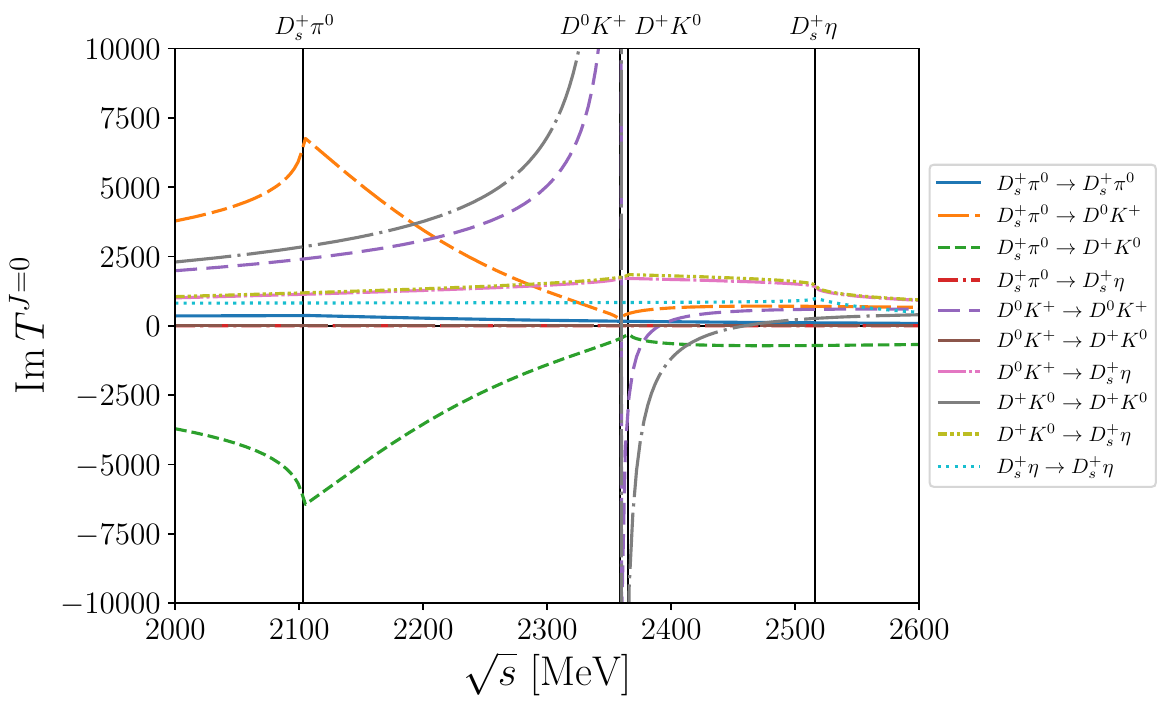}
    \includegraphics[scale=.45]{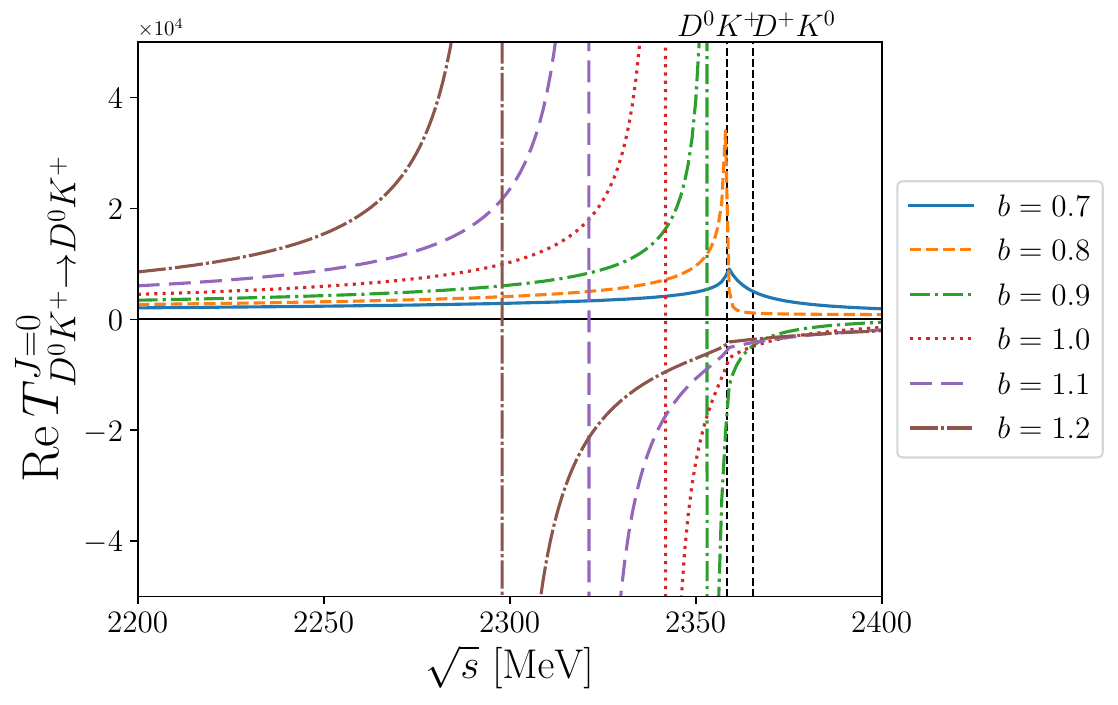}
    \caption{The real part of the single channel $T$-matrix elements
    are illustrated in the left panel, whereas in the right panel we
    show the effects of the $u$-channel diagrams in the single channel
    integral equation with parameter $b$ varied from 0.7 to 1.2.}
    \label{fig:3}
\end{figure} 
By solving the two-body integral equation with a single channel, we
can identify which channels give rise to resonant states. Each 
$T$-matrix element is depicted in the left panel of
Fig.~\ref{fig:3}. We see that the strong attraction in 
the $u$-channel exchange generates two poles in the $D^0K^+$ and
$D^+K^0$ channels. Though the kernel amplitudes for the $D_s^+\pi^0\to
DK$ and and $D_s^+\eta \to DK$ transitions also incorporate the
$u$-channel diagrams, no poles appear in the corresponding
$T$-matrix. Nevertheless, the $D_s^+\pi^0\to D^0K^+$ and
$D_s^+\pi^0\to D^+K^0$ contributions show a significant enhacement at
the $D_s^+\pi^0$ threshold energy. No pole structures were found in
other $T$ matrix elements.

To investigate the channel effects, we introduce a parameter $b$ to
serve as a tool to illustrate the coupling between different 
channels and to demonstrate the emergence of poles in the complex
plane, and multiply by it the coupling strength
$g_{DD^*\mathcal{M}}$. Setting $b$ to 0 turns off the corresponding channel,
allowing us to observe the behavior when that particular channel is
excluded. Conversely, setting $b$ to 1 represents the full
contribution of the transition matrix in the current work. In the
right panel of Fig.~ \ref{fig:3}, we depict the $u$-channel
dependence of the $T_{D^0K^+\to D^0K^+}^{J=0}$ with $b$ varied from
$b=0.7$ to $b=1.2$. Note that the pole, which arises from the strong
attraction in the $D^0K^+$ kernel amplitude, appears only if the value
of $b$ is approximately greater than 0.7. As anticipated, the energy 
corresponding to this pole increases as $b$ becomes larger.  

\begin{figure}[htp]
    \centering
    \includegraphics[scale=.55]{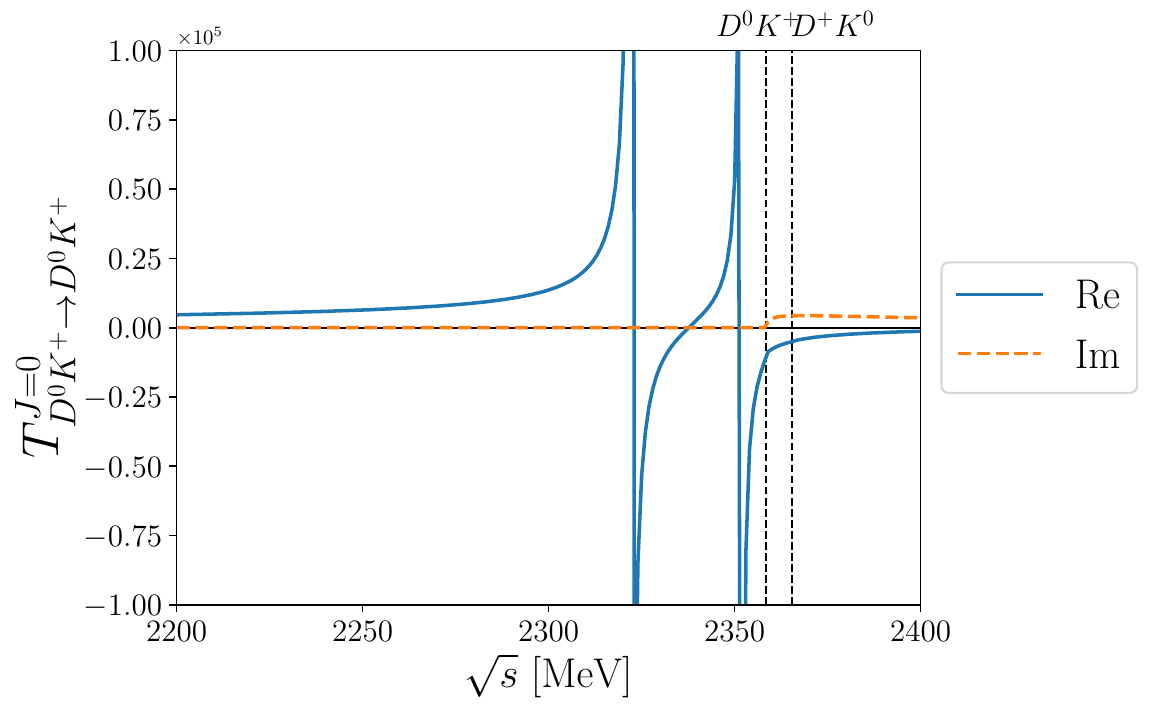}
    \caption{$J=0$ parital-wave transition amplitude for the $D^0K^+\to
      D^0K^+$ process when the $D^0K^+$ and $D^+K^0$ channels are
      coupled each other. The two poles lie below the $D^0K^+$
      threshold in the real axis.} 
    \label{fig:4}
\end{figure}
We now consider the S-wave transition amplitude for the $D^0K^+ \to
D^0K^+$ process. When the $D^0K^+$ channel is coupled to the $D^+K^0$
channel, the two poles still remain in the real $\sqrt{s}$ axis
below the $D^0K^+$ threshold at $\sqrt{s_R}=2322.04$ MeV and
$2351.82$ MeV, respectively, as depicted in Fig.~\ref{fig:4}. 
\begin{figure}[htp]
    \centering
    \includegraphics[scale=.46]{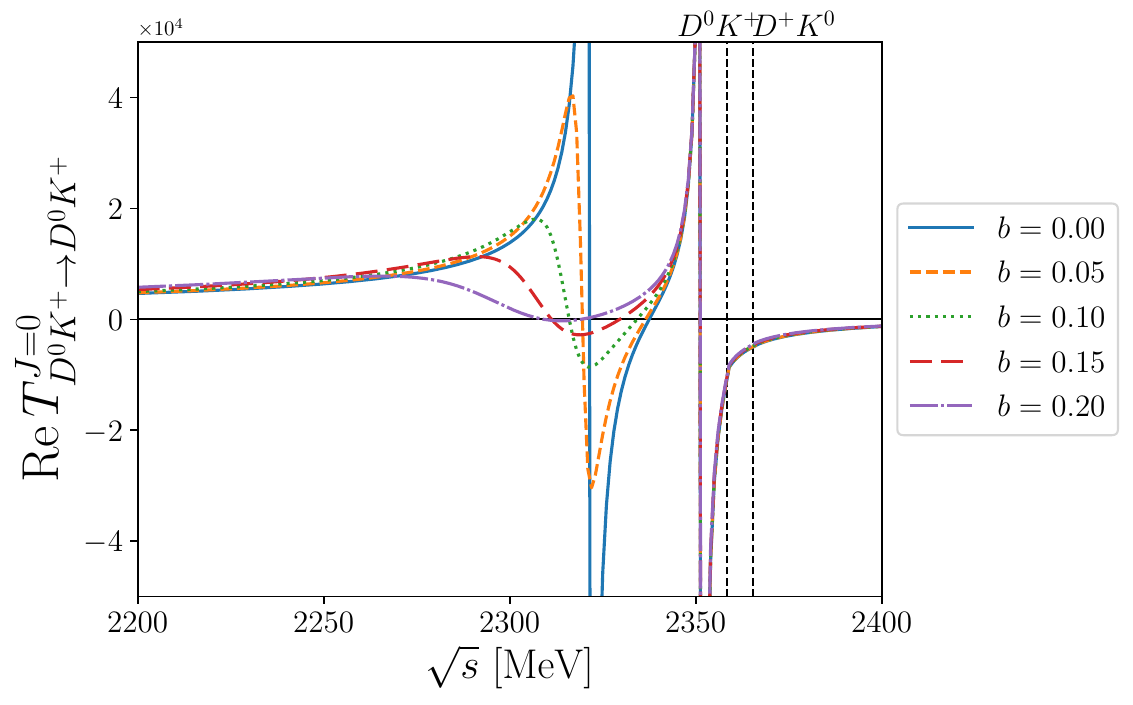}
    \includegraphics[scale=.46]{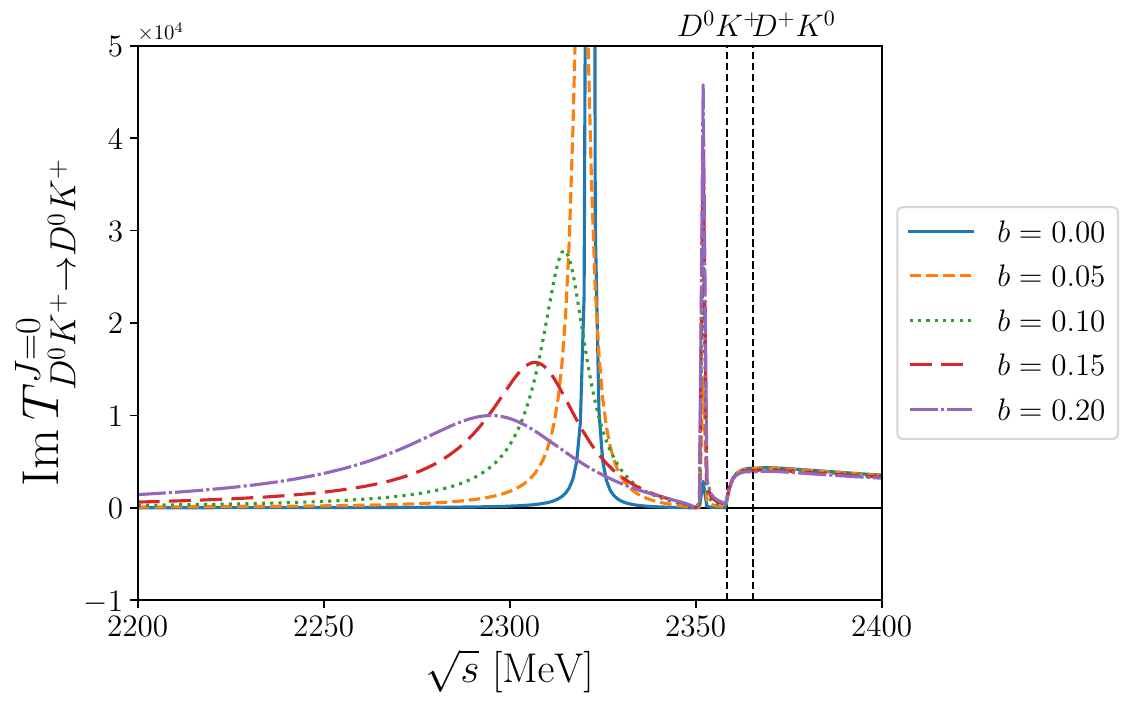}
    \caption{$D_s^+\pi^0$ channel coupled to the two $DK$
    channels. The real and imaginary parts of $T_{D^0K^+\to
    D^0K^+}$ are depicted in the left and right panel,
  respectively. $b$ varies from 0.00 to 0.20.}
    \label{fig:5}
\end{figure} 
However, once we couple it to one of the
non-diagonal amplitudes, only the first pole starts to undergo
changes. For example, let us couple the $D_s^+\pi^0$ amplitude to the 
$D^0K^+\to D^+K^0$ transition. We can scrutinize the coupled-channel
effects by varying the coupling strength of the two channels, which is
controlled by parameter $b$. As $b$ increases, the width of the first
pole is rapidly broadened, as shown in Fig.~\ref{fig:5}. When
$b$ reaches $0.69$, the deeper pole is already located at about
$\sqrt{s_R} \simeq 2110-i380$ MeV which is very close to the
$D_s^+\pi^0$ channel. Exceeding a value of $b=0.7$, this pole crosses
the $D_s^+\pi^0$ threshold energy and subsequently disappear from the
complex plane. By setting $b=1$, which corresponds to our final
results, the first pole completely disappears. Consequently, we
observe only the second pole that remains largely unaffected by the
$D_s^+\pi^0$ channel. 

\begin{figure}[htp]
    \centering
    \includegraphics[scale=.47]{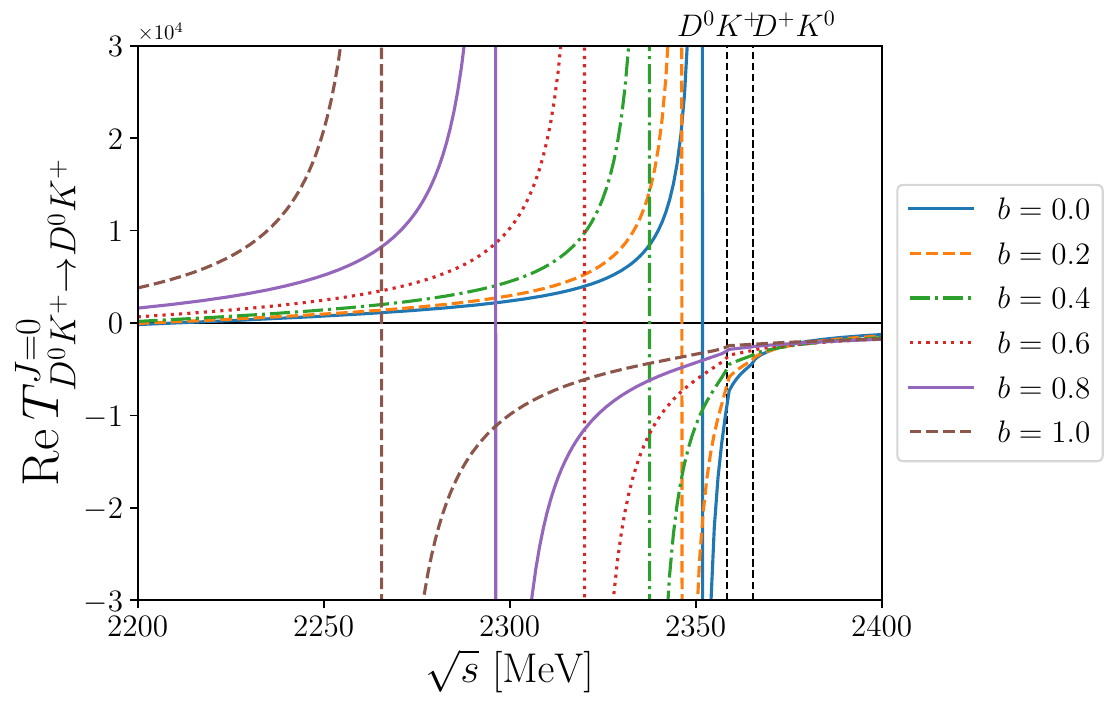}
    \includegraphics[scale=.47]{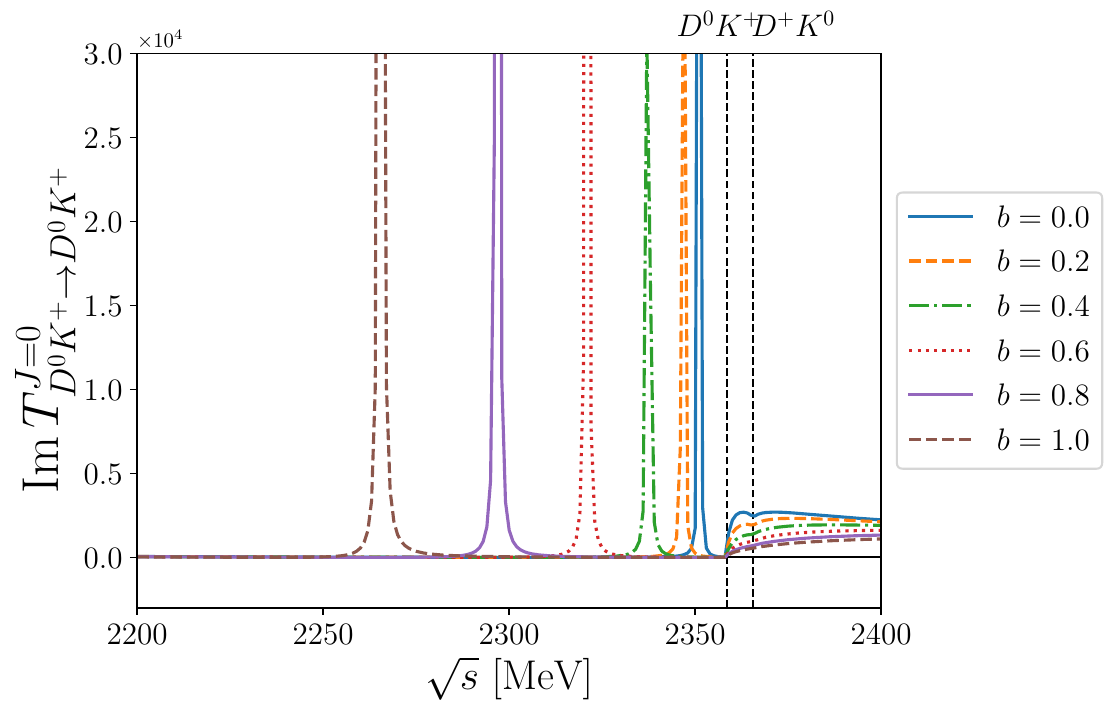}
    \caption{$J=0$ parital-wave transition amplitude coupled to the
      $D_s^+\eta$ channel with $b$ varied from 0.0 to 1.0. The real
      and imaginary parts of $T_{D^0K^+\to D^0K^+}$ are depicted in
      the left and right panel, respectively.}  
    \label{fig:6}
\end{figure} 
Next, we discuss how the $D_s^+\eta$ channel comes into play. 
If we couple $D_s^+\pi^0$, $DK$, and $D_s^+\eta$ channels together,
the second pole starts to move to the lower energy, which is
demonstrated in Fig.~\ref{fig:6}. When we introduce all
possible coupled channels, the pole is finally positioned at
approximately $\sqrt{s_R}=(2261.4-i0.072)$ MeV.

\begin{figure}[htp]
    \centering
    \includegraphics[scale=.6]{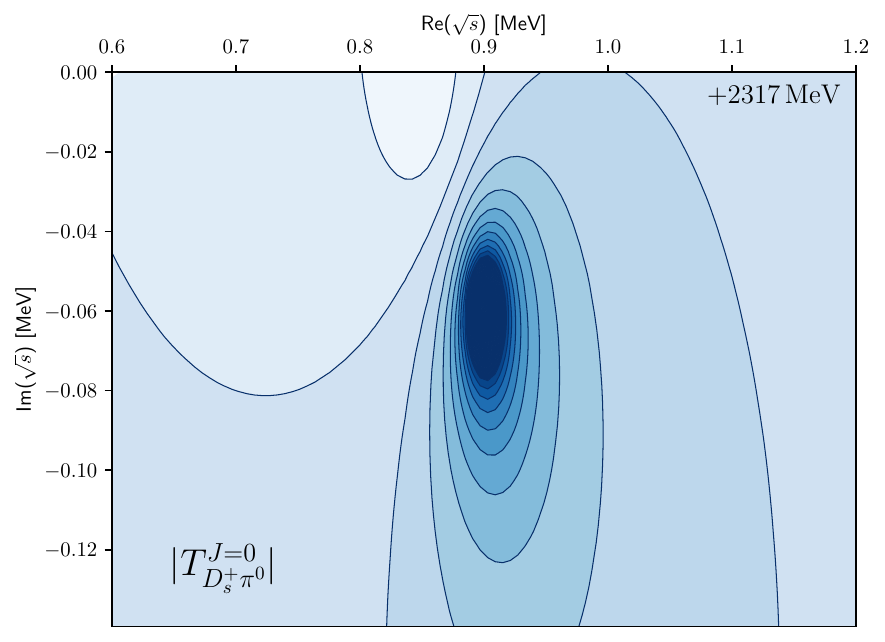}
    \caption{Contour plot of the $D_{s0}^*(2317)$. The pole
    position is predicted as $\sqrt{s_R} = (2317.90 - i0.0593)$ MeV.}
    \label{fig:7}
\end{figure}

The pole position on the real axis $\mathrm{Re}\,\sqrt{s_R} = 2261.4$
MeV still deviates from the experimental data for the mass of 
$D_{s0}^*(2317)$. Consequently,  it is necessary to adjust the model
parameters, with the uncertainties kept to be minimized.
We itemize how we proceed to fit the $D_{s0}^*(2317)$ mass: 
\begin{itemize}
  \item We can slightly change the cutoff masses listed in 
    Table~\ref{tab:2}. 
  \item While we still have one more free parameter, i.e.,
    $g_{\eta\eta\sigma}$, its effect is almost negligible. 
  \item We can also consider the effects of the explicit flavor SU(3)
    symmetry breaking, but find that they are also
    negligible.   
  \end{itemize}
Thus, we try to fit the cutoff masses to ensure 
that the resulting values align with our prescribed criterion
$\Lambda \simeq (m_{\mathrm{ex}} + 600)$ MeV as far as possible. The
cutoff masses for the exchanged light mesons, detailed in
Table~\ref{tab:2}, are maintained at $\Lambda = (m_{\mathrm{ex}} +
600)$ MeV in the $t$ channel, while slightly smaller values are chosen
for the exchanged heavy mesons in the $u$ channel, specifically
$\Lambda \approx (m_{\mathrm{ex}} + 540)$ MeV.  
Ultimately, this fitting process leads us to the physical pole
position situated in the second Riemann sheet: $\sqrt{s_R} = (2317.90
- i0.0593)$ MeV, visually depicted in Fig.~\ref{fig:7}. The figure 
illustrates the pole corresponding to the $D_{s0}^*(2317)$ meson in the
complex plane.
The pole for the $D_{s0}^*(2317)$ meson is located approximately 40 MeV
below the $DK$ thresholds, which indicates that the $D_{s0}^*(2317)$
may be identified as a $DK$ molecular state~\cite{Kolomeitsev:2003ac,
  Guo:2006fu, Lutz:2007sk}. 

To explore the effects of the coupled channels in more detail, we
analyze the channel couplings for the channels involved. 
The transition amplitudes in the vicinity of the pole position
$\sqrt{s_R}$ can be parameterized as
\begin{align} \label{}
T_{ab}(\sqrt{s}) = 4\pi\frac{ g_a g_b}{s-s_R}.
\end{align} 
The residue can be extracted by taking the limit $s\to s_R$ as
follows: 
\begin{align} \label{}
\mathcal{R}_{ab} = \lim_{s\to s_R}(s_R-s)T_{ab}/(4\pi).
\end{align}
The channel coupling is defined by 
\begin{align} \label{}
g_{a} = \sqrt{\mathcal{R}_{aa}},
\end{align}
which characterizes the coupling strength of the corresponding
resonance with a given channel $a$. We predict the four different
channel couplings as follows:
\begin{align} \label{}
&g_{D_s^+\pi^0} = 0.00967 + i0.05294\,\mathrm{GeV}, \;\;\; 
g_{D_s^+\eta} = -85.25 - i0.0610\,\mathrm{GeV}, \cr
&g_{D^0K^+} = -77.59 + i0.730\,\mathrm{GeV}, \;\;\;
g_{D^+K^0} = -80.16 - i0.905\,\mathrm{GeV}.
\end{align}
We can also define the corresponding coupling strengths: 
\begin{align} 
&g_1 = |g_{D_s^+\pi^0}| = 5.381\times 10^{-2}\,\mathrm{GeV}, \cr
&g_2 = |g_{D^0K^+}| = 77.59\,\mathrm{GeV}, \cr
&g_3 = |g_{D^+K^0}| = 80.17\,\mathrm{GeV}, \cr
&g_4 = |g_{D_s^+\eta}| = 85.25\,\mathrm{GeV}.
\label{eq:34}
\end{align}
Note that the channels corresponding to $g_i$ ($i=2,\,3,\,4$) are
not experimentally reachable by kinematics. The value of $g_1$ is the
predicted one, since all the parameters are fixed by reproducing the
mass of the $D_{s0}^*(2317)$ meson. Using the predicted value of
$g_1$, we determine the $D_{s0}^{*+}\to D_s^+\pi^0$ decay width: 
\begin{align} \label{}
\Gamma_{D_{s0}^{*+}\to D_s^+\pi^0} = 
\frac{g_1^2}{m_{D_{s0}^*}} \rho_{D_s^+\pi^0}(m_{D_{s0}^*}^2)
= \frac{g_1^2}{m_{D_{s0}^*}^2} \frac{p_\mathrm{cm}}{4\pi} =
13.86\,\mathrm{keV},
\end{align}
where $\rho_{R}(s)$ stands for the two-body phase space
factor and $p_\mathrm{cm}$ denotes the center of momentum
for the final states. The predicted value $\Gamma_{D_{s0}^{*+}\to
  D_s^+\pi^0} = 13.86\,\mathrm{keV}$ lies within 
the experimental upper limit of $3.8$ MeV. 

To scrutinize the reason for the very small value of the partial
width, we investgate two significant sources of the decay width. One
is the $\pi^0$-$\eta$ mixing effect discussed previously, while 
another sources are the mass differences between the neutral $D^0$
($K^0$) and the charged $D^+$ ($K^+$). If we neglect their mass 
differences, then we can specifically focus on the contribution of
$\pi^0$-$\eta$ mixing to the decay width. Interestingly, we observe
that the decay width becomes approximately doubled, i.e., $\mathrm{Im}
\sqrt{s_R} = -0.1070$ MeV, in comparison with the original 
results. This indicates that the 
$\pi^0-\eta$ mixing is destructively interfered by isospin effects
arising from the mass differences between the $D^0$ ($K^0$) and the
$D^+$ ($K^+$). Accordingly, the channel coupling $g_1$ becomes
$1.21 \times 10^{-1}$ GeV, resulting in a partial decay width of
$\Gamma_{D_{s0}^{*+} \to D_s^+ \pi^0} = 70.55$ keV. 

The coupling strengths convey information on which
channel is the strongest one. We conclude from the results in
Eq.~\eqref{eq:34} that the $DK$ channel is the most dominant one, and
the $D_s^+\eta$ is the second most significant one. In
Table~\ref{tab:3}, we compare the current results for the channel
couplings with those from other works. Interestingly, The values of
the $DK$ and $D_s^+\eta$ coupling strengths from the present work are
much larger than those from other works. 
\setlength{\tabcolsep}{5pt}
\renewcommand{\arraystretch}{1.5}
\begin{table}[htp]
\caption{Comparion with different models.} 
\label{tab:3}
\begin{tabular}{c| c c c c}
\hline
\hline
 & Present work  & Ref.~\cite{Fu:2021wde} 
 & Ref.~\cite{Lutz:2007sk} &
 Ref.~\cite{Gamermann:2006nm}\\
\hline 
$m_R$ [MeV] & $2317.90$ & $2318$ & $2317.6$ & $2317.25$ \\
$\Gamma_{D_{s0}^*}$ [keV] & $13.86$ & $132$ &
$140$ & - \\
$g_{D_s^+\pi^0}$ [GeV] & $5.381\times 10^{-3}$ & - & - & -
\\
$g_{D^0K^+}$ [GeV] & $77.59$ & $9.4$ & $7.579$ & $9.08$
\\
$g_{D^+K^0}$ [GeV] & $80.17$ & $9.4$ & $7.579$ & $9.08$
\\
$g_{D_s^+\eta}$ [GeV] & $85.25$ & $7.4$ & $5.795$ & $5.25$
\\[2pt]
\hline
\hline
\end{tabular}%
\end{table}

\begin{figure}[htp]
\centering
\includegraphics[scale=.6]{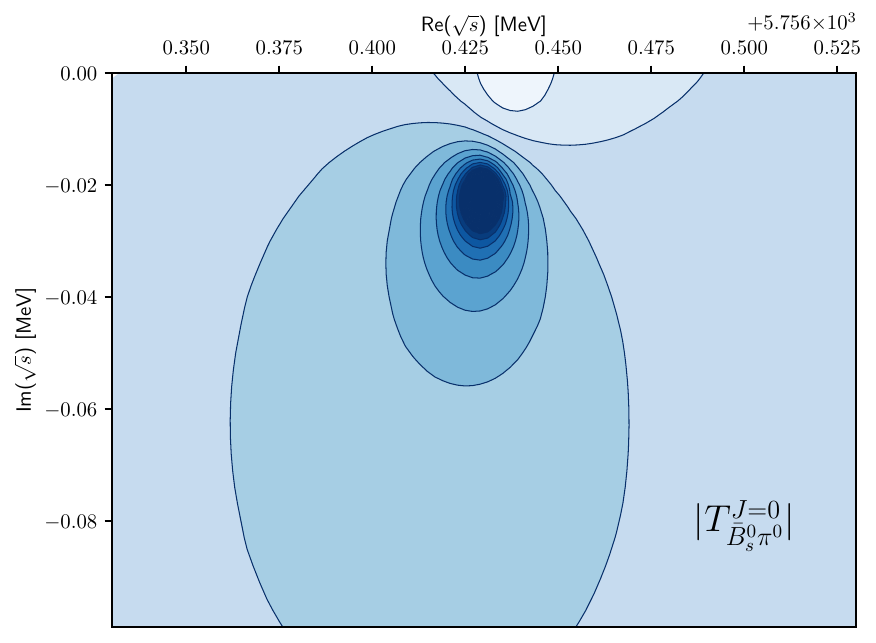}
\caption{Contour plot of the $\bar{B}_{s0}^*$ state. The
predicted pole position is $\sqrt{s_R} = (5756.43 
-i0.0215)$ MeV.}
\label{fig:8}
\end{figure}

In the same theoretical framework, our analysis expands to include the
scalar bottom-strange meson. Specifically, we employ the
coupled-channel formalism to delve into the $\bar{B}\mathcal{M}$
interaction with a positive strangeness, $S=+1$. It is crucial
to highlight a distinction from the charm sector; 
the charged $B_s^*$ meson is absent in the $u$-channel
process of the $B^-K^+$ channel. This means that the $u$-channel
contribution to the bottom sector becomes smaller than to the charm
sector. To compensate this absence, we prefer to use a lower value of
the cutoff mass for $B_s^{*0}$. So, we use the following values of 
$\Lambda_{B^*}=(m_{B^*} + 630)$ MeV and $\Lambda_{B_s^*}=(m_{B_s^*}
+390)$ MeV. Note that if we adhere to 600 MeV for the difference
between the mass of the exchanged particle and its cutoff value, then
the $B_{s0}^*$ would not emerge as a molecular state in the present
scheme. The pole for the $B_{s0}^*$ reappears in the second
Riemann sheet, positioned at $\sqrt{s_R} =(5756.43 - i0.0215)$ MeV,  
as visually represented in Fig.~\ref{fig:8}. It is noteworthy that the
total width of the $B_{s0}^*$ resonance, denoted as $\Gamma_{\bar{B}
{s0}^*}$, exhibits a comparable value to that of the $D_{s0}^*$
resonance. This observation is natural when considering that the
two-body phase space factor, $\rho_{\bar{B}_{s}^{0}\pi^0}(m_{\bar{B}_
{s0}^*}^2)$ for the $\bar{B}_{s0}^*$ resonance, is approximately two
times smaller than that of the $D_{s0}^*$. 

We exmaine the channel couplings for the bottom-strange meson as
follows: 
\begin{align} \label{}
&g_{\bar{B}_s^0\pi^0} = 0.0778 - i0.0675\,\mathrm{GeV},
\;\;\;
g_{\bar{B}_s^0\eta} = -133.81 + i5.11\,\mathrm{GeV}, \cr
&g_{B^-K^+} = -394.52 + i0.299\,\mathrm{GeV}, \;\;\;
g_{\bar{B}^0K^0} = -301.6 - i1.388\,\mathrm{GeV}.
\end{align}
The corresponding coupling strengths are given by 
\begin{align} \label{}
&g_1 = |g_{\bar{B}_s^0\pi^0}| = 0.103\,\mathrm{GeV}, \cr
&g_2 = |g_{B^-K^+}| = 301.64\,\mathrm{GeV}, \cr
&g_3 = |g_{\bar{B}^0K^0}| = 394.52\,\mathrm{GeV}, \cr
&g_4 = |g_{\bar{B}_s^0\eta}| = 133.9\,\mathrm{GeV}.
\end{align}
In the bottom sector, the $BK$ channel is the most dominant
one. As previously discussed, the strength of the $B^-K^+$
channel coupling is smaller than that of the $\bar{B}^0K^0$ channel
due to the absence of the $u$-channel diagram in the $B^-K^+$.
Furthurmore, the value of $g_4$ is smaller than that of $g_2$ and
$g_3$ by the same reason. Since we use smaller values of the cutoff
masses, we want to mention that the present finding does not guarantee
the existence of the $\bar{B}_{s0}^*$. However, if it is found by an
experiment, we can get a hint that the structure of $B_s^*$, which
brings about the main contribution, is different from other heavy
mesons considered in the current theoretical description, and the
$\bar{B}_{s0}^*$ is also understood as a molecular state.    

\section{Summary and conclusions}
In the current work, we aimed to investigate the $D_{s0}^*(2317)$,
studying $D_s^+\pi^0$ scattering within the off-shell coupled channel
formalism. We introduced four different coupled channels in the
charge basis: $D_s^+\pi^0$, $D^0K^+$, $D^+ K^0$ and $D_s^+\eta$ with 
positive strangeness $+1$. We first constructed the invariant kernel 
amplitudes based on the effective Lagrangians derived from heavy-quark
effective theory and hidden local symmetry. Having inserted the kernel
amplitudes into the Blankenbecler-Sugar-type coupled integral
equations, we evaluated the transition amplitudes. Since the
$D_{s0}^*(2317)$ is the scalar meson, we carried out the partial-wave
expansion to pick out the S-wave transition amplitude. We then
scrutinized the $DK$ channels how the pole appears
dynamically. When we only considered the $DK$ channels, two poles
emerged. Adding different channels one by one, the width of the first
pole became broadened. Once we included all the coupled channels, the
first pole disappeared, and at the same time the position of the
second pole started to move to lower energies. Having fitted the
cutoff masses, the pole is positioned at the physical mass of the
$D_{s0}^*(2317)$. The $D_{s0}^*\to D_s^+ \pi^0$ decay width was
predicted to be $13.86$ keV, which lies within the experimental upper
limit. The $DK$ channels were found to be the most dominant one.
We observed that the two sources of the decay width interfere
destructively with one another. This leads to the small value of the
partial decay width of $D_{s0}^{^+}\to D_s^+\pi^0$. The $D_{s0}^*$ is
positioned below the $DK$ threshold, indicating that it can be
interpreted as the $DK$ molecular state. On the other hand, the
generation of the $\bar{B}_{s0}^*$ requires the smaller value of the
cutoff mass due to the absence of the charged $B_s^*$ exchange
process. Thus, the existence of $\bar{B}_{s0}^*$ is less evident than
that of $D_{s0}^*$. 

\begin{acknowledgments}
We are very grateful to Samson Clymton for useful comments and
discussion. The present work was supported by Basic Science Research
Program through the National Research Foundation of Korea funded by
the Korean government (Ministry of Education, Science and Technology, 
MEST), Grant-No. 2021R1A2C2093368 and 2018R1A5A1025563. 
\end{acknowledgments}

 \appendix
 \section{SU(3) relations} \label{app:a1}
We provide the flavor SU(3) relations for the coupling constants for
the $HH\mathcal{M}$ and $HHV$ vertices.
 \begin{align} \label{}
 &g_{D_s^+K^+D^{*0}} = g_{D_s^+K^0D^{*+}} = g_{DD^*P}, \cr
 &g_{\pi^0D^0D^{*0}} = -g_{\pi^0D^+D^{*+}} 
 = g_{DD^*P}/\sqrt{2}, \cr
 &g_{\eta D^0D^{*0}} = g_{\eta D^+D^{*+}}
 = g_{DD^*P}/\sqrt{6}, \cr
 &g_{D_s^+D^0K^{*+}} = g_{D_s^+D^+K^{*0}} = g_{DDV}, \cr
 &g_{D^0D^0\rho^0} = g_{D^0D^0\omega} = g_{DDV}/\sqrt{2},
 \end{align}
 and
 \begin{align} \label{}
 &g_{\bar{B}_s^0K^+\bar{B}^{*-}} = g_{\bar{B}_s^0K^0\bar{B}^
 {*0}} = g_{BB^*P}, \cr
 &g_{\pi^0\bar{B}^0\bar{B}^{*0}} = -g_{\pi^0B^-B^{*-}} 
 = g_{BB^*P}/\sqrt{2}, \cr
 &g_{\eta \bar{B}^0B^{*0}} = g_{\eta B^-B^{*-}}
 = g_{DD^*P}/\sqrt{6}, \cr
 &g_{\bar{B}_s^0\bar{B}^0K^{*+}} = g_{\bar{B}_s^0B^-K^{*0}} =
 g_{BBV}, \cr
 &g_{\bar{B}^0\bar{B}^0\rho^0} = g_{\bar{B}^0\bar{B}^0\omega}
 = g_{BBV}/\sqrt{2}.
 \end{align}
 For the $\mathcal{M}\mathcal{M}\mathcal{M}$ and $\mathcal{M}
 \mathcal{M}V$ vertices, we have 
 \begin{align} \label{}
 &g_{\pi^0K^0K^{*0}} = -g_{\pi^0K^+K^{*+}} = g_{PPV},\cr
 &g_{\eta K^0K^{*0}} = g_{\eta K^+K^{*+}} = -\sqrt{3}g_
 {PPV},\cr
 &g_{K^+K^+\rho^0} = g_{K^0K^0\rho^0} =
 g_{K^+K^0\rho^-}/\sqrt{2}= g_{PPV}, \cr
 &g_{K^+K^+\omega} = -g_{K^0K^0\omega} = g_{PPV}.
 \end{align}

\bibliography{Dsubsj}
\bibliographystyle{apsrev4-1}

\end{document}